\documentstyle[twoside,fleqn,espcrc2,epsfig]{article}

\newcommand{\ifig}[1]{\mbox{\epsfig{file=#1,height=80mm,width=75mm}}}
\newcommand{\AC} {{\cal{A}}}
\newcommand{\VC} {{\cal{V}}}
\newcommand{\be}{\begin{equation}}
\newcommand{\ee}{\end{equation}}

\newcommand{\AmS}{{\protect\the\textfont2
A\kern-.1667em\lower.5ex\hbox{M}\kern-.125emS}}
\title{
On the Definition of Gauge Field Operators
in Lattice Gauge-Fixed Theories
       }

\author{L. Giusti\address{Scuola Normale Superiore, 
                P.zza dei Cavalieri 7, I-56100 Pisa Italy}
               \address{INFN, Sezione di Pisa, San Piero a Grado, I-56100
                      Pisa, Italy},
 M. L. Paciello\address{INFN, Sezione di Roma 1,
 P.le A. Moro 2, I-00185 Roma, Italy},
 S. Petrarca\thanks{Speaker at Lattice 98}$^{c~}$\address{Dipartimento di Fisica, Universit\`a di Roma "La
                     Sapienza"}%
        ,
B. Taglienti$^{c}$, M. Testa$^{c~d}$
                     }
       
\begin{document}

\begin{abstract}
We address the problem of defining  the gluon field
on the lattice
in terms of the natural link variables.
Different regularized definitions are shown, through non perturbative
numerical computation, to converge towards the same continuum renormalized
limit.
\end{abstract}

\maketitle
This talk, based on  ref. \cite{giusto} to which we refer for
details,  is divided in two parts. In the first one we will discuss the definition
of the gauge potential $A_ {\mu} (x)$ on the lattice and we will show
that  different definitions could have strong effects on 
gauge dependent quantities which are relevant in the  gauge-fixing procedure. 
Nevertheless we will show in the second part that 
different definitions
of the gluon field on the lattice give rise to Green's functions proportional
to each other at the non-perturbative level. This 
important feature is necessary in order to
garantee the uniqueness of the renormalized continuum operators.

The usual definition of the 4-potential in terms of the
links, $U_{\mu}$, which represent the fundamental dynamical gluon variables,
is given by
\be
{A}_{\mu} (x) \
\equiv \ {{( (U_{\mu} (x)) - (U_{\mu}^{\dagger} (x)))_{traceless}}
\over {2 i a g_0}}.  \label{eq:prima}
\ee
This definition is obtained taking an expansion in powers of $a$
of the relation $ U_{\mu}(x)\equiv \exp(i g_0 a A_{\mu} (x))$ that is
naively suggested by the interpretation of $U_{\mu}(x)$
as the lattice parallel transport operator and by
its formal expression in terms of
the "continuum" gauge field variables, $ A_{\mu}(x)$.
The definition given in eq.(\ref{eq:prima}) is not unique:
it cannot be preferred to any other definition with analogue properties
as, for instance:
\be
{A^{'}}_{\mu} (x) \
\equiv \ {{( (U_{\mu} (x))^2 - (U_{\mu}^{\dagger} (x))^2 )_{traceless}}
\over {4 i a g_0}},  \label{eq:seconda}
\ee
which in fact differs from  eq.(\ref{eq:prima})
by terms of $O(a^2)$ that formally go to zero as $a \rightarrow 0$.

\begin{figure}[htb]
\vspace{9pt}
\ifig{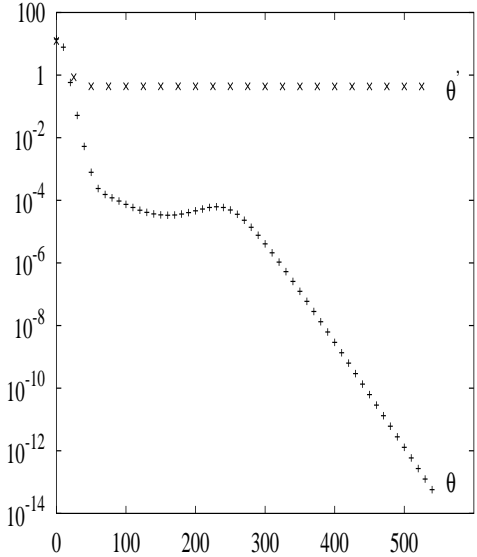}
\caption{\small{Typical behaviour of $\theta$ and $\theta^{'}$ vs gauge fixing sweeps at $\beta=6.0$
for a thermalized $SU(3)$ configuration $8^3\cdot 16$.}}
\label{fig:p60t}
\end{figure}

From the algorithmical point of view, however, the various definitions are
not interchangeable. In order to show this fact, 
let us suppose  to fix the Landau gauge
$\partial_{\mu} A_{\mu}$. First of all  one has to choose the functional
form of $A_{\mu}$ on the lattice in terms of the links and we
adopt the usual definition eq.(\ref{eq:prima}).
Then the gauge is fixed
applying a chain of gauge transformations
$U_{\mu}^{\Omega}(x) \equiv \Omega(x)
U_{\mu}(x) \Omega(x + \mu)^{\dagger}$  to a thermalized 
configuration until the control quantity
$\theta^{\Omega}\sim  \int d^4x( \partial_{\mu} A^{\Omega}_{\mu} )^{2}$
becomes very small, for example $\theta < 10^{-14}$.
Let us now  define $\theta^{'}$ with the
same functional form of $\theta$
but with $A_\mu$ replaced by $A^{'}_\mu$.
The values of $\theta$ and $\theta^{'}$,
are shown in Fig.\ref{fig:p60t}
for a typical thermalized
configuration, as functions of the lattice
sweeps of the numerical gauge-fixing algorithm.
As clearly seen
$\theta^{'}$ does not follow the same decreasing behaviour as $\theta$:
after an initial decrease, $\theta^{'}$ goes to a constant value, many orders of
magnitude higher than the corresponding value of $\theta$.
The marked difference between the two behaviours, already 
stressed in ref.\cite{giusti},
seems to cast doubts on the lattice gauge-fixing procedure
and on the correspondence among the continuum limits of gauge-dependent
operators having the same quantum numbers.
This discrepancy vanishes
if the comparison between two composite operators 
constructed in terms of $A$ and $A^{'}$ 
is done by checking  the values 
of the corresponding matrix elements
as must be done in field theory. Then on the lattice
one has to compare the average of the corresponding
matrix elements
taken on an ensemble of configurations.
The relation between the two lattice definitions $A_{\mu}(x)$
and $A^{'}_{\mu}(x)$ 
is of the form:
$
{A^{'}}_{\mu}(x)= A_{\mu}(x)+ a^2 W_\mu(x) 
$
where the $W_\mu(x)$ is a dimension $3$ unrenormalized operator
with the same quantum numbers of
$A_{\mu}(x)$.
The contribution of the operator $W $ to the relation between
$A$ and $A^{'}$ in the continuum limit can not be neglected
as it is shown from the following contruction of the renormalized
operator:
$
W^R_\mu(x) = Z_W(g_0,a\,\mu_{_R})(W_\mu(x)+ {1-C(g_0) \over a^2}
A_{\mu}(x)) 
$
($C$, as a consequence of the Callan-Symanzik equation,
can only depend on the bare coupling $g_0$\cite{prep})
so that
from the two last eqs. one obtains,
 up to terms truly of order $a^2$,
$
{A^{'}}_{\mu}(x)=C(g_0) A_{\mu}(x).
$
This operatorial relation implies on Green's functions that we in general:
\be
{{\langle \dots A'_{\mu}(x) \dots \rangle} \over {\langle \dots A_{\mu}(x)
\dots \rangle}}= C(g_0) \label{eq:green}
\ee 
We have numerically checked  eq.(\ref{eq:green})
by measuring on different $SU(3)$
lattices,
(see Table~\ref{tab:params}),
 in the Landau gauge with periodic
boundary conditions, a few interesting correlators which
are relevant to the
investigation of the QCD gluon sector.
The Landau gauge has been fixed in the standard way
\cite{mand_g,Davies} minimizing, for each
thermalized configuration,
the usual functional $F$:
$
F [U^{\Omega}] \equiv - \frac{1}{V\cdot T} 
Re \ Tr  \sum_{\mu} \sum_{x} U_{\mu}^{\Omega}(x).
$

In the following we will 
discuss these correlators:
\be
\langle {\AC}_0{\AC}_0\rangle (t) \equiv  \frac{1}{V^2}
\sum_{{\bf x},{\bf y}} Tr \langle  A_0({\bf x},t)A_0({\bf y},0) \rangle 
\label{eq:A0A0}
\ee
\be
\langle \AC_i\AC_i\rangle (t) \equiv  \frac{1}{3 V^2}  
 \sum_{i}\sum_{{\bf x},{\bf y}} Tr \langle  A_i({\bf x},t)A_i({\bf y},0)\rangle 
\label{eq:AiAi}
\ee
using both $A$ and $A^{'}$ as defined in eqs.(\ref{eq:prima}),
(\ref{eq:seconda}).
Here and in the following we define $\AC_{\mu}(t)=\sum_{\bf x} A_{\mu}({\bf x},t)$.
The correlation function $\langle {\AC}_0{\AC}_0\rangle (t)$, when
evaluated through $A_\mu(x)$, is
constant in $t$ configuration by configuration, in virtue of 
the Landau gauge condition which, together with periodic boundary
conditions, implies $\partial_0 \AC_0 = 0$.
The same should be true, on average, when $A^{'}$ is used.
The behaviours of these correlators 
(that we do not show here)
are well confirmed by our numerical simulations.
It is surprising  that also $\langle \AC^{'}_0\AC^{'}_0\rangle$
turns out to be constant configuration by configuration at the level of
$\sim5\%$, also because in this case 
the value of $\theta^{'}$
is different from zero on individual configurations as shown in Fig.~\ref{fig:p60t}.

\begin{figure}[htb]
\ifig{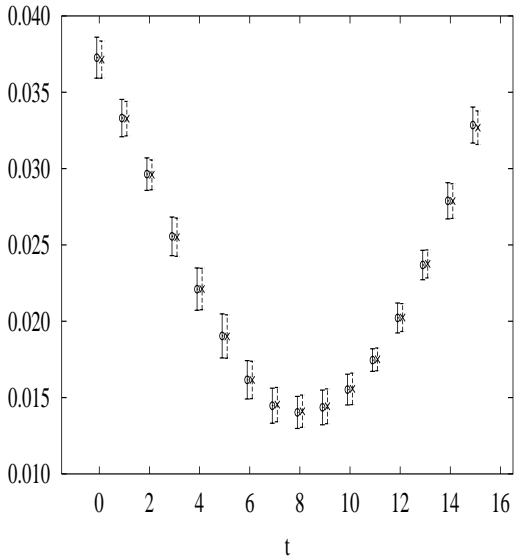}
\caption{\small{Comparison of the matrix elements 
of  $\langle\AC^{'}_i\AC^{'}_i\rangle(t)$ 
(crosses) and the rescaled $\langle\AC_i\AC_i\rangle~\cdot~C_i^2(g_0)$
(open circles) as function of time for 
a set of 50 thermalized $SU(3)$  configurations at $\beta=6.0$ with a 
volume $V\cdot T=8^3\cdot 16$ (run W60b). The data have been
slightly displaced in $t$ to help eye, the errors are jacknife.}}
\label{fig:amu}
\end{figure}

In Fig.~\ref{fig:amu} the Green function $\langle \AC^{'}_i\AC^{'}_i\rangle$
and the rescaled one
$C_i^2(g_0) \langle \AC_i\AC_i\rangle$ are reported
 for the run W60b.
The remarkable agreement between these two quantities
confirms the proportionality shown in eq. (\ref{eq:green})
(a triumph of field theory).

We have found that 
the proportionality factor, $C(g_0)$, may depend on the 
space-time direction
$\mu$.
 This fact is
 due to the breaking of cubic symmetry
and it could be a potential
source of systematic error in the non-perturbative evaluation of
renormalization constants on asymmetric lattices. 
In our simulations, as shown in Table~\ref{tab:params},
two of the lattices (W60b, W64),
have the time extension different from the spatial one,
so that we have a coefficient $C_0(g_0)^2$ relating 
$\langle {\AC}'_0{\AC}'_0 \rangle$
 to $\langle {\AC}_0{\AC}_0 \rangle$ and a
different one, $C_i(g_0)^2$, connecting $\langle {\AC}'_i{\AC}'_i \rangle$
 to $\langle {\AC}_i{\AC}_i \rangle$.
On the other hand, the values of $C_0(g_0)$ and $C_i(g_0)$
coincide, within the errors, for the symmetric lattices
and it is remarkable that the value of $C_0$ for W60a agrees within
the errors with the value of $C_i$ for W60b being the time extension 
of W60a equal to the spatial extension of W60b.

We are now ready to show why the discrepancy between
the values of $\theta$, relevant to control the gauge-fixing algorithm,
and the expectation values of $\theta^{'}$, is natural.
In fact  the definition of $\theta $  ($\theta^{'}$) is given by:
$
\theta = \frac{1}{V \cdot T} \
\sum_{ x} \theta( x) = \frac{1}{V \cdot T}
\ \sum_{x} Tr \ [ \Delta (x) \Delta^{\dagger} (x)]\;,
$
where:
$
\Delta( x) = \sum_{\mu} \ 
( A ({ x}) - A(x - \hat{\mu} ))\; 
$ and $A$  ($A^{'}$) is defined as in 
eq.(\ref{eq:prima})  ( eq.(\ref{eq:seconda})) without $a g_0$ to the denominator.
Then in the continuum variables
$
\theta ={ a^4 \over \VC }\int d^4x \ (\partial_\mu A_\mu(x))^2$
 where $\VC$ is the 4-volume in physical
units  (analogously  for  $\theta^{'}$).
Therefore, while $\theta$ vanishes configuration by configuration,
as a consequence of the gauge fixing, $\theta^{'}$ is proportional
to $(\partial_\mu A'_\mu)^2$, which has the vacuum quantum numbers and
mixes with the identity. The expectation value of
$(\partial_\mu A'_\mu)^2$, therefore, diverges as ${1 \over a^4}$
so that $\theta^{'}$ will stay finite, as $a \rightarrow 0$.

We believe that this discussion 
on the definition of gauge field operators
has a general validity and will
survive a more thorough treatment of the gauge-fixing problem.
 
\setlength{\tabcolsep}{.16pc}
\begin{table}
\begin{center} 
\begin{tabular}{||c|cccc||}
\hline\hline       
&W58&W60a&W60b&W64\\
\hline
$\beta$ & $5.8$       &  $6.0$      &     $6.0$    &   $6.4$      \\
\# Confs&  20         &  100        &     50       &   30         \\
Volume  &$6^3\times 6$&$8^3\times 8$&$8^3\times 16$&$8^3\times 16$\\
\hline
$C_i(g_0)$&    0.689(3)     &   0.729(1)  &   0.729(2)   &   0.757(2)   \\
$C_0(g_0)$&    0.690(7)     &   0.729(1)  &   0.750(1)   &   0.784(2)   \\
\hline
$a^{-1}$ & 1.333(6) & 1.94(5) & 1.94(5) & 3.62(4)  \\ 
\hline
\end{tabular}
\end{center}
\caption{\small{Summary of the lattice parameters used and relative
values of $C_0$ and $C_i$.
}}
\label{tab:params}
\end{table}
%

\end{document}